\documentclass[%
 reprint,
nofootinbib,
 amsmath,amssymb,
 aps,
]{revtex4-1}

\usepackage{graphicx}% Include figure files
\usepackage{dcolumn}% Align table columns on decimal point
\usepackage{bm}% bold math
\usepackage{xcolor}
\newcommand{\hslashslash}{%
  \raisebox{.85ex}{%
    \scalebox{.7}{%
      \rotatebox[origin=c]{18}{$-$}%
    }%
  }%
}
\newcommand{\bbar}{%
  {%
   \vphantom{b}%
   \kern0.09em
   \ooalign{\kern-0.16em\smash{\hslashslash}\hidewidth\cr$b$\cr}%
   \kern0.05em
  }%
}
\begin{document}

%\preprint{APS/123-QED}

\title{Escape from the Second Dimension: A Topological Distinction Between Edge and Screw Dislocations}% Force line breaks with \\

\author{Paul G. Severino}
 \email{paulsev@upenn.edu}%Lines break automatically or can be forced with \\
 \affiliation{
Department of Physics and Astronomy, University of Pennsylvania, 209 South 33rd St., Philadelphia, PA, 19104
}
\author{Randall D. Kamien}\email{kamien@upenn.edu}%
\affiliation{
Department of Physics and Astronomy, University of Pennsylvania, 209 South 33rd St., Philadelphia, PA, 19104
}

\date{\today}% It is always \today, today,

\begin{abstract}
Volterra's definition of dislocations in crystals distinguishes edge and screw defects geometrically, according to whether the Burgers vector is perpendicular or parallel to the defect.   Here, we demonstrate a distinction between screw and edge dislocations that enables a unified, purely topological means of classification.  Our construction relies on the construction of real or virtual disclination-line pairs at the core of the dislocation in a smectic and can be generalized to crystals with triply-periodic order.  The connection between  topology and geometry is exploited.
\end{abstract}

\maketitle

%\tableofcontents

Last year marked the 50th anniversary of de Gennes' celebrated analogy between smectic liquid crystals and superconductors  \cite{DEGENNES1972753}.  The relationship between the two systems has led to a greater understanding of smectic phases, prominently providing portent for the prediction of the twist grain boundary phase, an analogy to Abrikosov lattices in type II superconductors \cite{tgb_lubensky_renn}.  However, the structure of defects in smectics is profoundly different than that of the vortices in superconductors and superfluids.  Not only do smectics enjoy disclinations, defects with no analog in superconductors, but smectics also differ from superconductors by breaking translational symmetries  \cite{Mermin_RevMod_Defects}.  Note that vortex lines in superconductors have no special direction, but dislocations in smectics come in two types: edge and screw, defined via their geometry.  The standard way of viewing dislocations -- that of Volterra  \cite{Volterra1907} -- is much older than de Gennes' analogy.  The Volterra Process explicitly takes advantage of geometry by cutting, shifting, and gluing layers from a perfect crystal to create defects  \cite{chaikin_and_lubensky}.  The direction of the Burgers vector with respect to the defect naturally falls out of this procedure, affording edge and screw dislocations their usual distinction -- when the Burgers vector is parallel to the defect it is screw and when it is perpendicular it is edge.

While geometric constructions {\sl alla} Volterra are useful for visualizing dislocations and disclinations, classification of defects and their properties falls within the realm of topology.  However, it was recognized that in systems with broken translation order, the much-heralded homotopy theory approach to classify defects is inadequate: formally studying smectics as measured foliations, Po\'{e}naru proved that disclinations of charge $q>1$ cannot coexist with evenly spaced smectic layers in two dimensions \cite{poenaru}.  Disclinations in smectics thus do not form a group in the usual homotopy sense \cite{chen_goldstone,Mermin_RevMod_Defects}.   Dislocations, on the other hand, can be described by the homotopic approach which, indeed, predicts integer-valued Burgers charges.  Nevertheless, the fundamental group of the ground state manifold (GSM) for smectics and for crystals does not distinguish between edge and screw dislocations \cite{Mermin_RevMod_Defects}, leaving their distinction to Volterra's geometric classification.  In this paper we will resolve this by demonstrating that there is a topological distinction between screw and edge dislocations in smectics and, by extension, in all crystals.   This distinction between the local defect set and boundary conditions also serves to answer the question, ``after inserting one defect in a smectic, how do you decide whether the second defect is screw or edge?'' Should you use the local layer normal to define the Burgers displacement or the layer normal at infinity?  The topology of the density wave that forms the smectic resolves this issue as well.  We will also demonstrate that in the limit of a single defect in an otherwise perfect ground state, the geometric and topological determinations match.

Smectic liquid crystals are phases with a one-dimensional density modulation, $\rho({\bf x)} = \rho_0 +\delta\rho\cos\Phi({\bf x})$ where $\Phi({\bf x})$ determines the phase of the modulation\footnote{It should not be confused with the volume fraction in ``phase field'' models.}.  Density maxima, the level sets $\Phi=2\pi\mathbb{Z}$, define the smectic layers with a director field parallel to the layer normal ${\bf N}=\nabla\Phi/\vert\nabla\Phi\vert$.  A free energy density that leads to evenly spaced flat layers is, for instance, \cite{AvilesGiga}\footnote{The normalization is chosen so that if we set $\Phi=z-u({\bf x})$ then we arrive at the standard form of the free energy \cite{chaikin_and_lubensky}.}
\begin{equation}
{\cal F} = {B\over 8} \left[\left(\nabla\Phi\right)^2-1\right]^2 + {K\over 2}\left(\nabla^2\Phi\right)^2
\label{eq:fe}
\end{equation}
resulting in an absolute minimum when $\Phi={\bf k}\cdot{\bf x} + \phi$ where $|{\bf k}|=1$ chooses the direction of the ordering and $\phi$ is a constant offset.  
The density wave is invariant under $\Phi\rightarrow\Phi+2\pi$ and $\Phi\rightarrow-\Phi$, reflecting the discrete translational and rotational symmetries of the smectic.  Together, these invariances dictate that the manifold of ground states, represented by different directions of ${\bf k}$ and offsets $\phi$, is a Klein bottle \cite{trebin}.  
Dislocations are lines in three dimensions (or points in two dimensions) around which a Burgers circuit causes a displacement by some integer multiple of the lattice spacing.  This winding in the layer labeling causes $\Phi$ to be ill-defined on the dislocation.  Since the density must be well-defined, this implies that  $\delta\rho$ must vanish and so the smectic order melts at the defect core.  As noted in \cite{Disclination_Pairs,aspects_topology_smectics,Peierls_Nabarro}, this melting can be avoided by breaking the dislocation into a disclination pair with one disclination on a density maximum and the other on a density minimum.  Explicitly, a rotation of ${\bf k}$ by $\pi$ requires $\phi\rightarrow 2\pi n -\phi$ to preserve the symmetries of $\Phi$: a translation by $\delta$ is equivalent to a rotation by $\pi$ followed by a shift by $2\pi n -\delta$.  Representing the rotation by $\pi$ as $\mathcal{F}$ and a shift by $\delta$ as $\mathcal{S}_\delta$, we have $\mathcal{S}_\delta = \mathcal{F}^{-1}\mathcal{S}_{2\pi n-\delta} \mathcal{F}$ for all $n\in\mathbb{Z}$.  From this relation we have $\mathcal{S}_{2\pi} = \mathcal{S}_{\pi}\mathcal{F}^{-1}\mathcal{S}_{-\pi}\mathcal{F}$, that is, a dislocation that induces a shift by $2\pi$ and can be broken into a half shift, a rotation by $-\pi$, a half shift {\sl backwards}, and a rotation by $\pi$.  The net shift is $0$ but is replaced by two rotations, {\sl i.e.} disclinations.  
Whether this is observed in experiment is a matter of energetics -- the core could melt into a nematic to lower the overall energy.  However, from the topology of the far field we can always construct a pure disclination configuration that fills the sample.  

The disclination pairs are, however, different for an edge dislocation and a screw dislocation.  For edge dislocations in two dimensions, the classic construction in terms of a $+{1\over 2}/\!-\!{1\over 2}$ disclination dipole \cite{hexatic} and the resulting two dimensional texture can be extended into the third dimension perpendicular to the plane of the defects to create a three-dimensional edge dislocation line as shown in Fig.~1(Left).   The screw dislocation, on the other hand, only exists in three-dimensions.  In contrast to edge dislocations, cross sections of the screw core correspond to a pair of $+{1\over 2}$ disclinations.  Were the two disclinations parallel to each other, the requisite screw symmetry of the dislocation would be broken.  It follows that the two disclinations wind around each other as a pair of regular helices as shown in Fig.~1(Right).   From the perspective of three-dimensional nematics, the defect geometries of the screw and edge core are topologically equivalent: recall that disclinations of a nematic director restricted to the plane are characterized by $\pi_1(\mathbb{R}P^1)=\mathbb{Z}$ while a three-dimensional nematic has disclinations characterized by $\pi_1(\mathbb{R}P^2) = \mathbb{Z}_2$.  A pair of line disclinations in a nematic with cross-sectional geometry corresponding to two $+{1\over 2}$ disclinations, thus, {\sl has no charge} -- it escapes into the third dimension \cite{escapeinto}. 

\begin{figure}
    \centering
    \includegraphics[scale=.22]{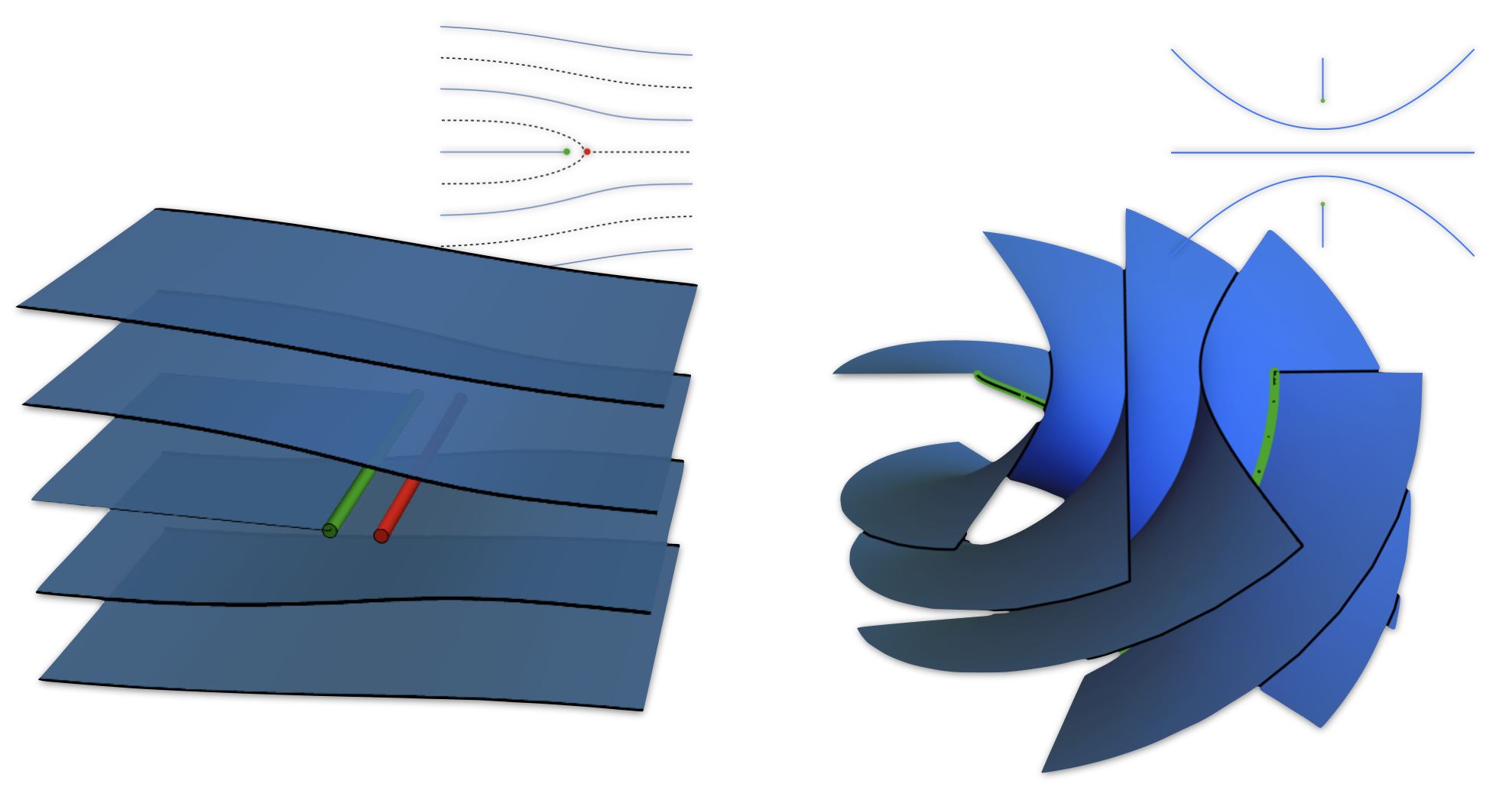}
    \caption{\label{screw_edge_fig}{(Fig.~1. (Left) Edge dislocation core in three dimensions decomposed as a +1/2 disclination on a density maximum and
a -1/2 on a density minimum. (Right) Screw dislocation core
as a double helix of +1/2 disclinations. Insets show cross
sections perpendicular to the dislocation.}}
    \label{fig:GSM}
\end{figure}

 In the nematic, this smooth escape and the related transition from a $+{1\over 2}$ geometry to a $-{1\over 2}$ geometry requires the director to twist -- a distortion that is impossible for a smectic!   Nonetheless, the transition from one dislocation core to the other can be achieved at a single point at which two extra layers attach to one of the +1/2 disclinations.  The location of the transition can slide freely along the dislocation without melting the smectic. 
At this point the Burgers' charge is unchanged, the winding of the phase field $\Phi$ is unchanged at large distances, but the disclination geometry changes.  
 
It is only because the smectic cannot transition smoothly from one disclination geometry to the other that we can exploit the dichotomy of their cores.  From this, it follows that we can distinguish a screw dislocation from an edge dislocation by examining the local texture of the layer normal around the disclination pair of which it comprises.  There is no need to reference a special direction for the Burgers charge, and, indeed, there is no real sense in which it is a vector in $\mathbb{R}^3$ -- the phase field $\Phi$ does not transform under rotations.  This is often obscured when one expands around a particular ground state so that $\Phi=z-u({\bf x})$ where the layers are normal to $\hat z$ and where $u({\bf x})$ is the standard Eulerian displacement field for a smectic\cite{chaikin_and_lubensky}.  In this case the transformation of $z$ requires a concomitant transformation in $u({\bf x})$.  From a Lagrangian point of view, where we parameterize deformation of the ground state in terms of displacements from a fiducial ground state, it is far more compelling to view the Burgers charge as a vector, especially in the case of three-dimensional crystals (to which we will turn our attention in the following).  The {\sl Volterra construction} relies on this Lagrangian form of elasticity and therein lies the need to distinguish edge and screw dislocations by {\sl global geometry.} 
 
 Two- and three-dimensional crystals can also be described via phase fields: a $d$-dimensional crystal is created via $\rho({\bf x}) = \rho_0 +\delta\rho\left[\sum_i^d \cos(\Phi_i)\right]$ so that the lattice points sit at the maxima of all the density waves.  Dislocations correspond to winding in one of the $\Phi_i$ and, in turn, this dislocation can be split, {\sl mutatis mutandi}, into disclinations associate with that phase field.  The remaining phase fields continue to provide a simple, periodic wave in the other directions, creating a dislocation with a Burgers charge in the $i^{\rm th}$ phase field.  More generally, a dislocation would be characterized by $d$ dislocation charges, one for each phase field.  In the ground state, $\Phi_i = {\bf G}_i\cdot{\bf x}+\phi_i$ where ${\bf G}_i$ are the reciprocal lattice vectors and $\phi_i$ set the origin of the crystal.  The topological charges associated with each $\Phi_i$ are $2\pi n_i$ with $n_i\in\mathbb{Z}$  and can be converted into the Burgers vector ${\bf b}=\sum_i n_i {\bf e_i}$ where the ${\bf e}_i$ are the basis vectors with ${\bf G}_i\cdot{\bf e_j} = 2\pi\delta_{ij}$.  
 
 From here, the discussion of splitting dislocations into disclination pairs which maintain separate density minima and maxima proceeds as with the smectic.  It is essential to note that the disclinations are in the phase fields $\Phi_i$ and are {\sl not} necessarily the geometric disclinations that the Volterra construction creates.  In the case of a smectic, they coincide but, in general the phase disclinations and geometric disclinations are distinct \cite{Peierls_Nabarro}.  Again, we can detect where a screw and edge dislocation meet by looking at the point transition from a $+{1\over 2}$ geometry to a $-{1\over 2}$ geometry.  In an otherwise perfect ground state, the standard distinction between edge and screw (Burgers vector parallel or perpendicular to the dislocation line) provides the same classification as the topological character of their cores.  What about adding a defect to a distorted ground state that is otherwise free of topological defects?  If the distortions vanish at the boundary then we can define asymptotic basis vectors ${\bf e}_i$ from which to specify the Burgers vector.  But a dislocation in the bulk need not remain a straight line -- indeed, any smooth deformation of a screw or edge dislocations preserves the topological characterization of their cores. Indeed, any diffeomorphism of the sample preserves the distinction between screw and edge because the transition from one to the other happens at a single point -- the point can move but the defect retains its character on either side of it.  Expanding the Burgers vector in the asymptotic conflicts with the standard geometric classification of screw versus edge.

To make this discussion concrete, we focus on the screw dislocation in a smectic.  A screw dislocations of Burgers displacement $b$ has the form
    $\Phi_{\text{screw}}=z-\frac{b}{2\pi}\arctan\left( y/x\right)$
at large distances.   The Euler-Lagrange equation for (\ref{eq:fe}) is:
\begin{equation}
0=\frac{1}{2}B \nabla\cdot\left[\nabla\Phi\left(\vert\nabla\Phi\vert^2-1\right)\right] - K\nabla^2\nabla^2\Phi
\end{equation}
Because $\Phi_{\text{screw}}$ is harmonic in $x$ and $y$ and the gradient of the compression (radial) is orthogonal to the layer normal (azimuthal and along $\hat z$), it satisfies the necessary conditions for a minimum.   However, all layers of a helicoid meet at the center to become one surface, yielding infinite compression energy at the screw defect core.  Rather than melting the smectic in this region, one may construct the core from a disclination pair as shown in Fig.~1(Right).  The construction starts with a central helicoid upon which subsequent layers are built by pushing off a constant distance along the layer normal.  By construction, this creates a family of evenly spaced layers with zero compression.  However, just as with evolutes in two-dimensional curves, this family of surfaces will develop line-like cusps on either side of the initial helicoid, the first pair forming a double helix at a radius $\rho_0$.  These first cusps are precisely the location of the $+{1\over 2}$ disclination geometry.  Thus a cylinder of radius $\rho_0$ can be constructed with zero compression and be used for the core of screw dislocation.  As shown in Fig.~1(Right), the primordial helicoid extends to the outside of the core and additional helicoids can be attached, as necessary, at $\rho_0$, yielding an outer solution.  Note that the cusps could occur at density maxima or density minima and the separation between them can be related directly to the Burgers charge of the defect \cite{screwone,screwtwo}.  

How is this different than the construction of a three-dimensional edge dislocation?  In an edge dislocation the pair of disclination lines are parallel and, in perpendicular cross section, reduces to the two-dimensional pair of point disclinations.  In cross section, however, a screw dislocation appears to be a pair of $+{1\over 2}$ disclinations, suggesting a total disclination charge of $+1$ in the system.  But far from the dislocation core the helicoid reduces to flat layers, as required by the smectic boundary conditions.  This apparent inconsistency is not present in the edge dislocation, which satisfies the boundary conditions by naturally having a disclination charge neutral core.  Without the ability to twist, the geometry of the helicoidal smectic layers must allow the disclinations of the screw core to relax.  To clarify the importance of the geometry, first consider the nematic director defined by $\nabla\Phi_{\text{screw}}$ at $z=0$:

\begin{equation}
   {\bf n}=\frac{\nabla\Phi_{\text{screw}}}{|\nabla\Phi_{\text{screw}}|}=\frac{1}{\sqrt{r^2+\bbar^2}}\left(-\bbar\sin\theta,\bbar\cos\theta,r  \right)
\end{equation}
where $r$, $\theta$, and $z$ are the usual cylindrical co\"ordinates and $\bbar\equiv b/(2\pi)$.

This form of $\bf n$ appears to have a charge $+1$ disclination at the origin.  Looking down the screw axis of the helicoid yields normals that wind by $2\pi$ around the center, hence the reason two $+1/2$ disclinations were required to build the core.  However, in the smectic phase the director is not defined homogeneously throughout the sample but instead only normal to the layers.  Rather than taking a projection for constant $z$, one only defines ${\bf n}||{\bf N}$ on a surface of constant $\Phi$.  For instance, taking $\Phi_{\text{screw}}=0$, {\sl i.e.} $\theta=(2\pi/b)z$ yields instead

\begin{equation}
    {\bf N}=\frac{1}{\sqrt{r^2+\bbar^2}}\left[-{\bbar}\sin\left(z/\bbar\right), {\bbar}\cos\left(z/\bbar\right),r   \right]
\end{equation}
at the point on the two-dimensional surface parameterized by $r$ and $\theta$, ${\cal P}=(r\cos\theta,r\sin\theta,\bbar\theta)$.

As $r\rightarrow 0$ we see that while the layer normal does rotate, it is still single valued at each fixed $z$ slice, as required.  The rotation of the normal by $2\pi$ is accompanied by a concomitant change in $z$.  The helicoid is simply connected, so in order to measure the disclination charge with a {\sl closed} loop, we must rely on the density wave, which is well-defined everywhere, to fill in the remaining space with ``virtual layers'' between the density maxima.  Only then do we see the nematic disclination on every constant $z$ slice.  However, as $r\rightarrow\infty$, ${\bf N}\rightarrow\hat z$ and ${\bf n}\rightarrow \hat z$.  At constant height $z$ there is no disclination charge at large $r$, despite the charge at small $r$ -- is disclination charge not conserved?  There is no contradiction since, in three dimensional nematics, a disclination line with geometric winding $+1$ can escape into the third dimension.  

In the smectic we see that 
the mechanism for relieving the apparent disclination charge and converting it into the Burgers displacement is the Gaussian curvature of the helicoid.  Though the Gaussian curvature of a two-dimensional manifold interacts with the defects of director fields lying in the local tangent plane \cite{lubenskymac}, here the situation is quite different; since the director is always normal to the surface, the director itself defines the surface.  The fact that in the smectic-A phase the director and layer normal are one in the same is the key to connecting topological defects to Gaussian curvature: it allows us to construct a Gauss map for disclinations in the director field.  Recall that the Gauss map is defined by taking the unit normal to a surface and mapping it to the center of a unit sphere.  The collection of normals to a patch of the original surface sweeps out a corresponding {\sl signed} area on the unit sphere and the Jacobian that converts area on the manifold to area on the sphere is the Gaussian curvature \cite{RevModPhys_Geometry}.

\begin{figure}
    \centering
    \includegraphics[scale=.3]{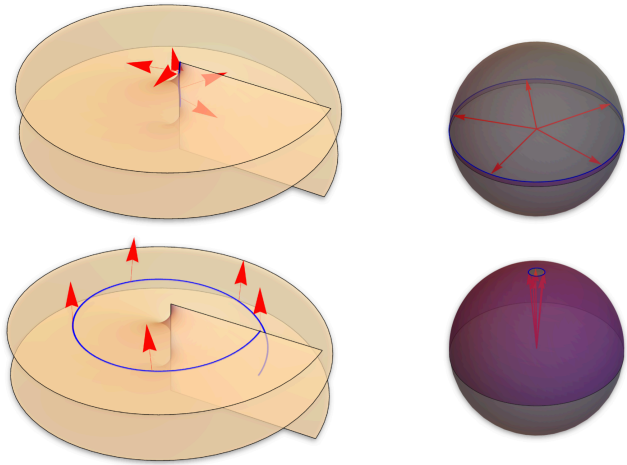}
    \caption{\label{gauss_map_figure}{The Gauss map on the helicoid for contours at varying distances from the core.  (Top) Near the core the geometric texture of the director winds by $2\pi$, and sweeps out the equator on the unit sphere.   (Bottom) Far from the core the director and normals align along $\hat{z}$ in accordance with boundary conditions.}}
    \label{fig:gauss_map_figure}
\end{figure}

With this in mind, consider a texture with the geometry of a $+1$ disclination line in a three-dimensional director field.  The charge can be measured by a circle taken around the disclination.  If the nematic director measured along this circle defined the normal to a smectic layer, we could use the director to to define a Gauss map to the unit sphere.  This measuring circuit may move between level sets while enclosing the defect, just like the measuring loop for the screw core.  By definition, the director along the measuring circuit will be mapped to the equator of the unit sphere.  Imposing the boundary conditions that the normals all point along $\hat{z}$ at infinity, a contour stretched out to infinity must map to a point on the unit sphere (the North pole): as the contour is stretched from the original measuring circuit to infinity, the Gauss map sweeps out area on the unit sphere and it follows that the surfaces must have Gaussian curvature.  Indeed, by checking the relative orientation of the areas we see that the surfaces must have a {\sl negative} integrated Gaussian curvature.  This is exactly what happens in the case of the screw dislocation, where the Gaussian curvature of the helicoid, $K=-\bbar^2/(r^2+\bbar^2)^2$
allows for the equatorial loop on the unit sphere to be moved up along to a point infinitely far from the helicoid core (Figure \ref{fig:gauss_map_figure}).  

It is enlightening to consider the saddle-splay term in the nematic
\begin{equation}
F_{\text{ss}} = -K_{24} \int d^3\!x \,\nabla\cdot{\bf A}
\end{equation}
where ${\bf A}= {\bf n}\left(\nabla\cdot{\bf n}\right)-\left({\bf n}\cdot\nabla\right){\bf n}$.  Because it is a total divergence it contributes only at boundaries, which include the cores of defects.  While we have shown that disclinations can be arranged at dislocation cores so that the smectic need not melt, the disclinations themselves require vanishing of the nematic order since ${\bf n}$ becomes undefined.  The saddle-splay integrates to the surface bounding the melted region.  Note that for an $m$-fold, two-dimensional disclination, ${\bf n}=[\cos m\theta,\sin m\theta,0]$ (with $m\in{1\over 2}\mathbb{Z}$) we find that 
${\bf A} = -m\hat r/r$ -- it has a nonzero divergence at the origin and is singular when $m\ne 0$.  Indeed, integrating around the inner boundary of the nematic (the disclination) in the $xy$-plane we find a total saddle splay of $2\pi m$.  What is this quantity?  Recalling that the saddle-splay $\nabla\cdot {\bf A}=2K$ is twice the Gaussian curvature of the surface.   The integrated curvature of a $+1$ dislocation or half-helicoid (that ends at the origin) is precisely $\pi$ and can be calculated directly from the Gaussian curvature, from the divergence theorem,  or as half of the upper hemisphere on the Gauss map.  Note that the illustration in Fig. \ref{fig:gauss_map_figure} displays the escape for a full helicoid that has no boundary at the origin.  Indeed the, saddle-splay vector, ${\bf A}$ reveals both the apparent disclination charge at the origin and the negative curvature required to let it escape from the second dimension!  This vector quantity is a new measure of the smectic complexion and allows us to distinguish edge from screw dislocations.  The core of screw dislocations needs to be constructed out of two disclinations with a $+1/2$ geometry.   This disclination ``escapes from the second dimension'' by following the normal to a family of helicoids.  Alternately, in cross section an edge dislocation has no net disclination charge and that is reflected by the fact that edge dislocations do not generate net Gaussian curvature in smectic layers.    Again, to generalize to three-dimensional crystals it is only required to examine each phase field separately as discussed above.  All the geometry and topology of the level sets follows the discussion for smectics.

It is also interesting to study other surfaces from the perspective of Gaussian curvature and disclination charge.  The intertwining of the disclinations to create a dislocation is essential.   One could consider concentric cylinders built from a pair of parallel $+1/2$ disclinations connected transversely along a layer -- the standard $+1$ disclination line in a three-dimensional smectic.  Now, building up smectic surfaces results in concentric cylinders which have vanishing Gaussian curvature (though nonvanishing mean curvature).   We could distort the locations of the disclination lines and instead create concentric surfaces resembling the catenoid.  In this case, though the layer normal relaxes to be parallel to $\hat z$, a cusp is formed in the concentric surfaces forming a new disclination loop, a geometry identical to that in the inner region of a toroidal focal conic domain. 

Our analysis brings about a more nuanced view of screw and edge dislocations both in smectics and in crystals.  The standard geometric approach distinguishes edge and screw by the direction of the Burgers vector while our analysis allows for a completely topological interpretation of the distinction.  This interpretation is essential when considering smooth deformations of crystals with embedded dislocations since the local geometry can deviate from the crystalline axes specified by the boundary conditions.  Whether further topological information regarding the entangling of dislocations and disclinations can be characterized is an open question.

We thank M. Cates for a penetrating question that clarified our results.  P.G.S. was supported by an NSF Graduate Fellowship.  R.D.K. acknowledges the hospitality of the Institute for Theoretical Physics at Utrecht University under the auspices of the Kramers Chair of Theoretical Physics.  This work was supported by a Simons Investigator grant from the Simons Foundation to R.D.K.

\providecommand{\noopsort}[1]{}\providecommand{\singleletter}[1]{#1}%
%

%\nocite{*}

%\bibliography{paper}% Produces the bibliography via BibTeX.

\end{document}